\newcommand{\gev}{\, {\rm GeV}}
\newcommand{\beq}{\begin{equation}}
\newcommand{\eeq}{\end{equation}}
\newcommand{\bea}{\begin{eqnarray}}
\newcommand{\eea}{\end{eqnarray}}

\documentclass[aps,prl,reprint,twocolumn,preprintnumbers,floatfix,nofootinbib]{revtex4-1}

\usepackage{graphicx}
\usepackage{epstopdf}
\usepackage{mathrsfs}
\usepackage{amssymb}
\usepackage{verbatim}
\usepackage{color}
\usepackage[dvipsnames]{xcolor}
\usepackage{multirow}
\usepackage{amsmath}
\usepackage{subfig}
 \usepackage{slashed} 
\usepackage{float}
\usepackage[normalem]{ulem}
\usepackage{sidecap}
\usepackage{hyperref}

\def\c{\chi}					
\def\xa{\text{\tiny{$X_1$}}}
\def\xna{\text{\tiny{$X_N$}}}
\def\zp{\text{$Z'$}}

\def\dm{\text{\tiny{DM}}}
\def\sm{\text{\tiny{SM}}}

\def\rh{\text{\tiny{RH}}}
\def\max{\text{\tiny{MAX}}}
\def\id{\text{\tiny{ID}}}

\def\a{\alpha}
\def\b{\beta}
\def\g{\gamma}
\def\s{\sigma}
\def\r{\rho}

\def\m{\mu}
\def\n{\nu}
\def\e{\epsilon}

\def\G{\Gamma}
\def\L{\Lambda}
\def\O{\Omega}

\def\to{\rightarrow}

\def\Lg{\mathcal{L}}

\usepackage[parfill]{parskip}
\begin{document}\sloppy 

\preprint{LPT--Orsay 18-14}
\preprint{}

\vspace*{1mm}

\title{Freezing-in dark matter through a heavy invisible $Z'$}
\author{Gautam Bhattacharyya$^{a}$}
\email{gautam.bhattacharyya@saha.ac.in}
\author{Ma\'ira Dutra$^{b}$}
\email{maira.dutra@th.u-psud.fr}
\author{Yann Mambrini$^{b}$}
\email{yann.mambrini@th.u-psud.fr}
\author{Mathias Pierre$^{b}$}
\email{mathias.pierre@th.u-psud.fr}
\vspace{0.5cm}
\affiliation{
${}^a $ Saha Institute of Nuclear Physics,
HBNI, 1/AF Bidhan Nagar, Kolkata 700064, India}
\affiliation{
${}^b$ 
Laboratoire de Physique Th\'eorique (UMR8627), CNRS, Universit\'e Paris-Sud, Universit\'e Paris-Saclay, 91405 Orsay, France}

\begin{abstract} 
We demonstrate that in a class of the $U(1)'$ extension of the Standard Model (SM), under which all the
Standard Model matter fields are uncharged and the additional neutral gauge boson $Z'$ couples to a set of
heavy nonstandard fermions, dark matter (DM) production mediated by $Z'$ can proceed through the
generation of generalized Chern-Simons (GCS) couplings. The origin of the GCS terms is intimately
connected to the cancellation of gauge anomalies. We show that the DM production cross section triggered by GCS couplings is sufficient even for an intermediate scale $Z'$ . A large range of DM and Z masses is then
allowed for reasonably high reheating temperature ($T_\rh \gtrsim 10^{10}~\text{GeV}$). This type of scenario opens up a new paradigm for unified models. We also study the UV completion of such effective field theory constructions, augmenting it by a heavy fermionic spectrum. The latter, when integrated out, generates the
GCS-like terms and provides a new portal to the dark sector. The presence of a number of derivative
couplings in the GCS-like operators induces a high temperature dependence to the DM production rate. The
mechanism has novel consequences and leads to a new reheating dependence of the relic abundance.
 \end{abstract}

\maketitle

\setcounter{equation}{0}

\section{I. Introduction}

In spite of a lot of speculations about its origin, dark matter (DM)
still remains an enigma, and the best we can do is to assume that it
has a particle physics origin in the domain of natural extensions of
the Standard Model (SM). However, the twin pressure of the clear
existence of DM in the energy budget of the Universe \cite{planck} and
simultaneously the lack of any DM signal in direct detection
experiments XENON \cite{XENON}, LUX \cite {LUX}, and PANDAX
\cite{PANDAX} pushes the limits on weakly interacting massive
particles (WIMPs) toward unnatural corners of the parameter
space. The simplest extensions as Higgs portal \cite{Higgsportal},
$Z$-portal \cite{Zportal}, or even $Z'$ portal \cite{Zpportal} are now
severely constrained (for a review on WIMP searches and supporting
models, see Ref.~\cite{Arcadi:2017kky}). This scenario motivates the
assumption that the interactions between the dark and visible sectors
are even weaker, leading to an out-of-equilibrium production of feebly
interacting massive particles (FIMPs) \cite{fimp} (see Ref.~\cite{Bernal:2017kxu} for a review). Alternative generation mechanisms
of a weakly interacting dark sector by direct thermal production at
the reheating temperatures are discussed in Ref.~\cite{alternative}.

On the other hand, theoretical considerations ranging from neutrino
mass generation mechanisms to grand unified theories (GUT), as well as
inflation, reheating, leptogenesis, or Higgs stability, all hint
toward the existence of an intermediate scale between $10^{10}$ and
$10^{16}$~GeV. To interpret the absence of DM signals,
instead of invoking unnaturally weak
DM-SM couplings, one
could explain its secluded nature by suppressions arising from the
high-mass scale of the mediators involved in its interaction with the
thermal bath. Concrete realizations for the high-energy physics origin
of the freeze-in mechanism involve Planck suppressed portals that
could be embedded in quantum theories of gravity
\cite{mambrinilast,Garny:2015sjg,PIDM}, the left-right symmetric model
\cite{Biswas:2018aib} and $Z'$ mediators in the $SO(10)$ framework
\cite{Mambrini:2013iaa,2loops}.

In usual $Z'$-mediated constructions, the SM particles are charged
under the new gauge group $U(1)'$. An interesting question, therefore,
is to ask if the DM production processes are still efficient even if
the SM is uncharged under $U(1)'$. To generate the effective
interaction of the associated gauge boson $Z'$ with the SM fields, we
would need a set of nonstandard fermions charged under $U(1)'$ as well
as under the SM gauge group(s). Such set-up is quite common in string
constructions, or in $E_6$ models. In this case, effective
interactions of the type represented by the Lagrangian \beq
{\cal L} \supset \lambda ~ \epsilon^{\mu \nu \rho \sigma} Z'_\mu A_\nu
F_{\rho \sigma} \, ,
\label{Eq:term}
\eeq where $F_{\rho\sigma} = \partial_\rho A_\sigma - \partial_\sigma
A_\rho$, arise from diagrams leading to anomalies (\`a la
Green-Schwarz in string models or Peccei-Quinn in the presence of
axionic couplings).  The gauge boson $A$ in the above expression may
be the SM hypercharge gauge boson or could even be any other
nonstandard $U(1)$ gauge boson.

Such terms, characterized by the presence of three gauge bosons and
one derivative, are dubbed generalized Chern-Simons (GCS) terms
\cite{GCS} (see also Ref.~\cite{anomaly-zpr} for a general discussion of
anomaly-free $Z'$ models). This operator can be generated at the
dimension-4 level as a low-energy effective term by integrating out a
set of heavy fermions. The effective coupling $\lambda$ is independent
of the heavy fermion masses, and hence its effect does not decouple by
increasing those fermion masses.  The underlying dynamics behind this
apparent nondecoupling is that this term is gauge noninvariant, but
its gauge variation cancels some triangle anomalies from some lighter
fermions still persisting in the theory. On the other hand, when we
integrate out a complete anomaly-free set of heavy fermions having
couplings with $Z'$ and $A$, similar but higher-dimensional terms with
three gauge bosons and more derivatives are generated, {\em albeit}
suppressed by the heavy fermion mass scale. We will indistinctly refer
to those also as GCS terms as in Eq.~(\ref{Eq:term}). Such terms have
already been studied as a connection between the SM field content and
dark sectors in thermal~\cite{CSDM} and
nonthermal~\cite{Farzan:2014foo} dark matter production mechanisms.

Now, let us suppose that $Z'$ couples to our DM candidate, leaving open
the possibilities that this candidate could be a fermion or a
vectorial boson (Abelian or non-Abelian). If the DM is a vectorial
boson, we also assume that it couples to $Z'$ through GCS interactions
generated possibly by a different set of heavy fermions. Notice that, even if the SM matter fields are {\em not} charged under $U(1)'$, $Z'$
would still be produced in the early Universe through the {\em freeze-in}
mechanism \cite{fimp}, thanks to the GCS terms, which would then decay
to DM particle $\chi$ : (SM) (SM) $\rightarrow Z' \rightarrow \chi
\chi$. However, the DM production rate is doubly suppressed, both by
the mass scale of heavy fermions in loop generating the GCS couplings
as well as by the mass of the virtual $Z'$ exchanged in the
process. In fact, through a large temperature dependence, this rate is
highly sensitive to the reheating phase, especially when considered
noninstantaneous, as shown in Refs.~\cite{Giudice:2000ex, Garcia:2017tuj}
(and Ref.~\cite{gravitino} for gravitino dark matter). Moreover, allowing
for vectorial dark matter (Abelian or non-Abelian) brings in higher-derivative couplings in the GCS operators, thereby inducing a
significantly high temperature sensitivity to the DM production rate.

The paper is organized as follows. We first describe the model under
consideration. Then, we compute the relic abundance from the freeze-in
process in the early thermal bath, taking into account
noninstantaneous reheating. We then construct models for UV
completion that would naturally lead to our framework. We finally
conclude, highlighting the new aspects that emerged from our analysis.

\section{II. Our model}

The two portals that connect the DM to the SM sector are $(i)$ a $Z'$
gauge boson of the $U(1)'$ group and $(ii)$ a set of heavy fermions
charged under $U(1)'$ as well as the SM gauge group(s). Depending on
the nature of the couplings of those heavy fermions with the SM gauge
bosons and the $Z'$, three possibilities emerge

\begin{itemize}
\item The heavy fermions are vectorlike with respect to SM gauge
  bosons as well as $U(1)'$. In this case, no GCS terms are generated, as in the absence of chiral couplings, any potential operator that
  contains three gauge bosons (and derivatives) vanishes identically
  \cite{Dudas:2013sia}.

\item The heavy fermions are chiral under $U(1)'$ but do {\it not}
  form an anomaly-free set. Then, GCS terms at dimension-4 level are
  generated in the effective action of which the gauge variations exactly
  cancel the anomaly \cite{GCS}, yielding

\beq {\cal L}= \alpha ~\theta' \text{Tr}[F^\sm F^\sm] + \beta
~\epsilon^{\mu \nu \rho \sigma} Z'_\mu A^\sm_\nu F^\sm_{\rho \sigma}
\, , 
\eeq with $A^\sm$ and $F^\sm$ being the hypercharge SM gauge
boson and associated field strength, respectively, and $\theta'$ being the
would-be Goldstone boson of $U(1)'$. Notice the absence of any
suppression coming from the masses of heavy fermions (generically
denoted by $M$) that are integrated out in generating the effective
couplings $\alpha$ and $\beta$. This apparent nondecoupling signifies
that the set-up is not anomaly free. These constructions are common in
string-inspired models (Green-Schwarz mechanism) and lead to
interesting phenomenological consequences (e.g., advocated to justify
monochromatic spectral lines originating from the dark sector
\cite{line}).

\item The heavy fermions form an anomaly-free set. They are chiral
  under $U(1)'$ but vectorlike with respect to the SM gauge group. In
  this case, several gauge-invariant combinations can be written. The
  complete list can be found in Ref.~\cite{Dudas:2013sia}. However, a lot
  of them\footnote{For instance, terms of the type
    $(1/M^2)\epsilon^{\mu \nu \rho \sigma} Z'_\mu (H^\dag D_\nu H)
    F^\text{SM}_{\rho \sigma}$ do not contribute to CS-like couplings
    above the electroweak phase transition. } either are not
  generated by triangle loops or vanish when the SM gauge bosons
  are on shell. In the limit where the field $\Phi$ that breaks the
  extra $U(1)'$ is much heavier than its vacuum expectation value
  (VEV), $V$, which controls the $Z'$ mass, the effective theory
  exhibits only the axionic (longitudinal) component ($a$) of the
  field as $\Phi = \frac{V + \phi}{\sqrt{2}}\exp(i a/V)$. Defining the
  dimensionless axion $\theta' \equiv \frac{a}{V}$, the only relevant
  coupling can then be extracted \cite{Dudas:2013sia} from a gauge
  invariant Lagrangian, given by
\beq {\cal L} = \frac{1}{M^2}
  \partial^\alpha {\cal D}_\alpha \theta'~ \epsilon^{\mu \nu \rho
    \sigma} \mathrm{Tr}[F^\sm_{\mu \nu} F^\sm_{\rho
      \sigma}] + V^2|{\cal D} \theta'|^2
\label{Eq:lagrangian1}
\eeq
with ${\cal D}_\alpha \theta' \equiv \partial_\alpha \theta' -
\frac{\tilde g}{2} Z'_\alpha$, $\tilde g$ being the gauge coupling
associated to the extra $U(1)'$.  Throughout the present work, we
shall consider this particular set-up, even if our results can be
applied to a general class of GCS couplings just by a
redefinition of parameters.
\end{itemize}

In the unitary gauge, the term related to the $Z'$-SM-SM vertex can be
extracted from Eq.~(\ref{Eq:lagrangian1}) as
\beq
{\cal L} = \frac{\tilde g}{M^2}\partial^\alpha Z'_\alpha \epsilon^{\mu
  \nu \rho \sigma} \partial_\mu A^a_\nu \partial_\rho A^a_\sigma \, , 
\label{Eq:lagrangian2}
\eeq
where $A^a$ are the SM gauge bosons. From now on, without any loss of
generality, we consider the gluons as the gauge bosons appearing in
Eq.~(\ref{Eq:lagrangian2}) and define $\frac{1}{\Lambda^2} \equiv
\frac{\tilde g}{ M^2}$, as the results would be exactly the same with
electroweak gauge bosons, just by rescaling the couplings. We consider
the heavy fermions generating the GCS couplings to be charged under
$SU(3)_C$ so that they dominate the production
process\footnote{Considering fermions without hypercharge $Y$ leads to
  a simplification as kinetic mixing of the type $\delta ~Z'_{\mu \nu}
  B^{\mu \nu}$ can be avoided. Effects of such mixing have been
  extensively studied in the literature \cite{kineticmix}}. With
this approach, the relevant Lagrangian would then read
\cite{Dudas:2013sia}
\begin{equation}
\Lg_\text{eff} = \frac{1}{ \Lambda^2} \partial^\a Z'_\a \e^{\m \n \r
  \s}\text{Tr}[ G^a_{\m \n} G^a_{\r \s}] + \Lg_\dm^i \, , 
\label{Eq:leff}
\end{equation}
where ${\cal L}^i_\dm$ represents the interactions between the $Z'$
and the DM candidate, which can be fermionic ($\c$), or vectorial of
Abelian ($X_1$) or non-Abelian ($X_N$) types. The respective
Lagrangians are given by\footnote{Only axial coupling is present for
  the fermionic dark matter. The derivative $\partial^\a$ before
  $Z'_\a$ in Eq.~(\ref{Eq:lagrangian2}) ensures that the vector
  coupling does not contribute in a $GG \to \c\c$ process.}
\begin{equation}
\Lg_\dm^\c = \alpha ~\bar{\c}\g^\m \g_5 \c Z'_{\mu},
\end{equation}
\begin{equation}
\Lg_\dm^{\xa} = \b ~ \e_{\m\n\r\s} Z'^\m X_1^\n X_1^{\r\s},
\end{equation}
and
\begin{equation}\label{Eq:zpxn}
\Lg_\dm^{\xna} = \g ~\partial^\a Z'_\a \e_{\m \n \r \s}\text{Tr}[ X_N^{\m \n} X_N^{\r \s}].
\end{equation}

\section{III. Results}

The evolution of dark matter number density $n_\dm$ is governed by the Boltzmann equation
\beq \label{dndt}
\frac{\text{d} n_\dm}{\text{d}t} = -3 H(T) n_\dm + R(T),  
\eeq
where $H \left( T \right)$ is the Hubble expansion rate and $R(T)$ is
the temperature-dependent interaction rate. In the regime in which the
abundance of dark matter is much smaller than the abundance of
particles in the thermal bath, the backreaction term in the rate
(dark matter producing standard particles) may be neglected, which is
usually the case in the freeze-in mechanism\footnote{The correct
  amount of dark matter is generated in a regime in which $n_\dm \ll
  n_\sm$, since $\frac{\O_\dm^0 h^2}{0.12} \sim
  \frac{Y_\dm}{10^{-10}}\frac{m_\dm}{\gev}$ and $Y_\dm \propto
  n_\dm/n_\sm$.}.

\begin{figure}[h!]
\centering
\includegraphics[scale=0.3]{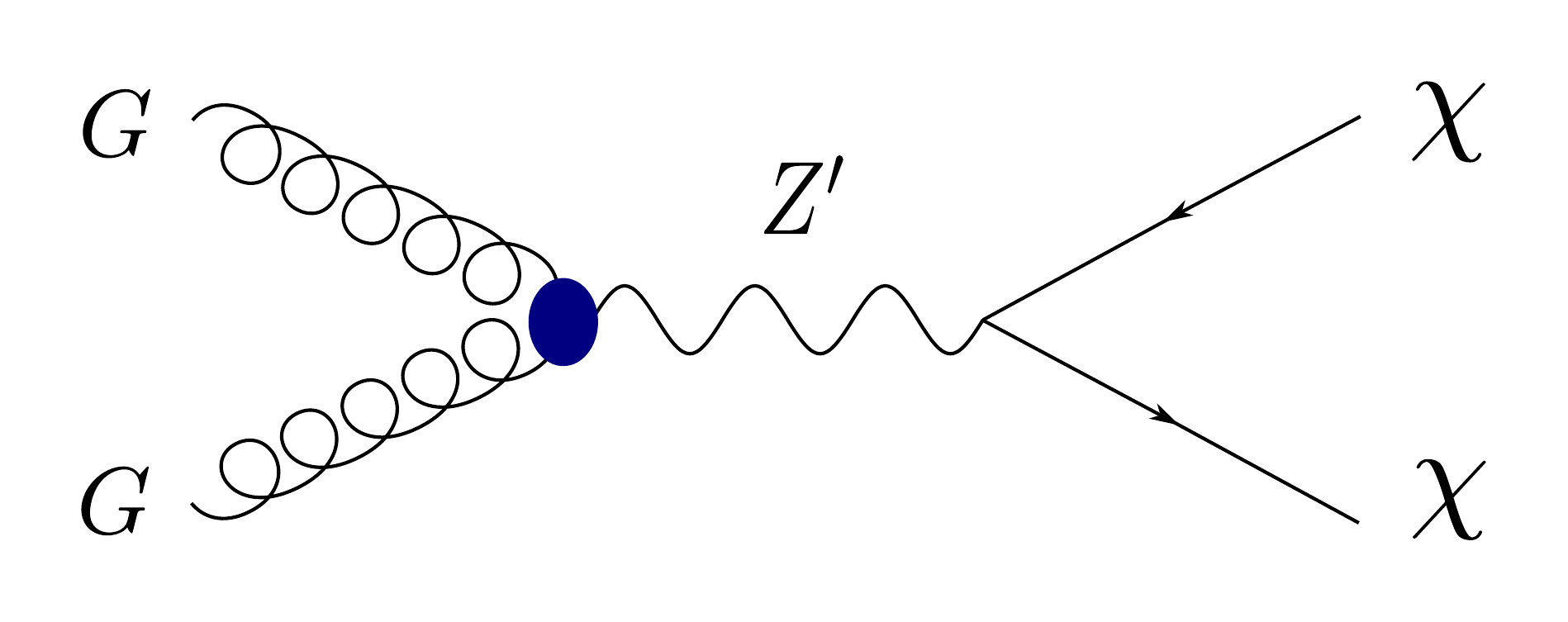}
\caption{\em \small Production of dark matter through gluon fusion in the early Universe}
\label{Fig:annihilation}
\end{figure}

In our set-up, the freeze-in occurs through the process depicted in
Fig.~\ref{Fig:annihilation}. For a 1 + 2 $\rightarrow$ 3 + 4 process,
the rate can be written as $R(T)= n_\text{sm}^2 \langle \sigma v
\rangle$, where $n_\text{sm}$ is the number density of the SM species and
$\langle \sigma v \rangle$ is the thermal averaged production
cross section. For a dark matter particle $i$, the rate reads
\beq R(T) = \int f_1 f_2 \frac{E_1 E_2 \text{d}E_1 \text{d}E_2
  ~\text{d}\cos \theta_{12}}{1024 \pi^6} \int |{\cal M}|_i^2 \text{d}
\Omega_{13} \,,
\label{Eq:rt}
\eeq

with $E_{1,2}$ and $f_{1,2}$ as the energy and the distribution
function of the initial SM particles, $\theta_{12}$ as the angle
between them, and $d \Omega_{13}$ as the solid angle between particles 1 and 3.

For the fermionic dark matter case,
\begin{equation}\begin{split}
\int |\mathcal{M}|_\c^2 ~\text{d}\O_{13} &= 2^{10} \pi
~\frac{\alpha^2}{\Lambda^4} \frac{m_\c^2 }{M_\zp^4}
\frac{s^3(s-M_\zp^2)^2}{(s-M_\zp^2)^2+M_\zp^2 \Gamma_\zp^2}\\ &
\approx 2^{10}\pi ~\frac{\a^2}{\Lambda^4} \frac{m_\c^2}{M_\zp^4}s^3
\label{Eq:m1}
\end{split}
\end{equation}
For the Abelian dark matter case,
\begin{equation}\begin{split}
\int |\mathcal{M}|_{\xa}^2 ~\text{d}\O_{13} &= 2^{10} \pi ~
\frac{\b^2}{\Lambda^4} \frac{s^3}{M_\zp^4} \frac{ (s-4m_{\xa}^2)
  (s-M_\zp^2)^2}{(s-M_\zp^2)^2+M_\zp^2 \Gamma_\zp^2} \\ & \approx
2^{10}\pi ~ \frac{\beta^2}{\Lambda^4} \frac{1}{M_\zp^4}s^4
\label{Eq:m2}
\end{split}
\end{equation}
For the non-Abelian dark matter case,
\begin{equation}\begin{split}
\int  |\mathcal{M}|_{\xna}^2 \text{d}\O_{13}&= 2^{12} \pi~
\frac{\g^2}{\Lambda^4} \frac{s^5}{M_\zp^4} \frac{
(s-4m_{\xna}^2)
(s-M_\zp^2)^2}{(s-M_\zp^2)^2+M_\zp^2 \Gamma_\zp^2}
\\ & \approx 2^{12}\pi ~ 
\frac{\g^2}{\Lambda^4}\frac{1}{M_\zp^4}s^6 
\label{Eq:m3}
\end{split}
\end{equation}

Above, $\Gamma_\zp$ is the total width of $Z'$ (see the Appendix for
details); $s$ is the center-of-mass energy squared; and $m_\c, m_\xa,
m_\xna$, and $M_\zp$ are the three types of dark matter and $Z'$
masses, respectively. Note that we recover the Landau-Yang effect in
the above expressions, though the pole enhancement studied in Ref.~\cite{mambrinilast} is not present in our case.  Note also that the
vectorial nature of the mediator has specific characteristics that we
do not observe for other type of mediators. Importantly, we notice
that once the pole is reached ($s = M_\zp^2$), the production rate
vanishes exactly -- see Eqs.~(\ref{Eq:m1}), (\ref{Eq:m2}), and (\ref{Eq:m3}). This is expected following the Landau-Yang theorem,
which states that a massive spin-1 particle cannot decay into two
massless spin-1 fields.  This behavior is opposite of the traditional
freeze-in scenario in which, on the contrary, the majority of dark matter
is produced when the temperature of the thermal bath reaches $T \sim
M_\zp$ \cite{mambrinilast,Blennow:2013jba}.

\begin{figure}[t]
\centering
\includegraphics[scale=0.5]{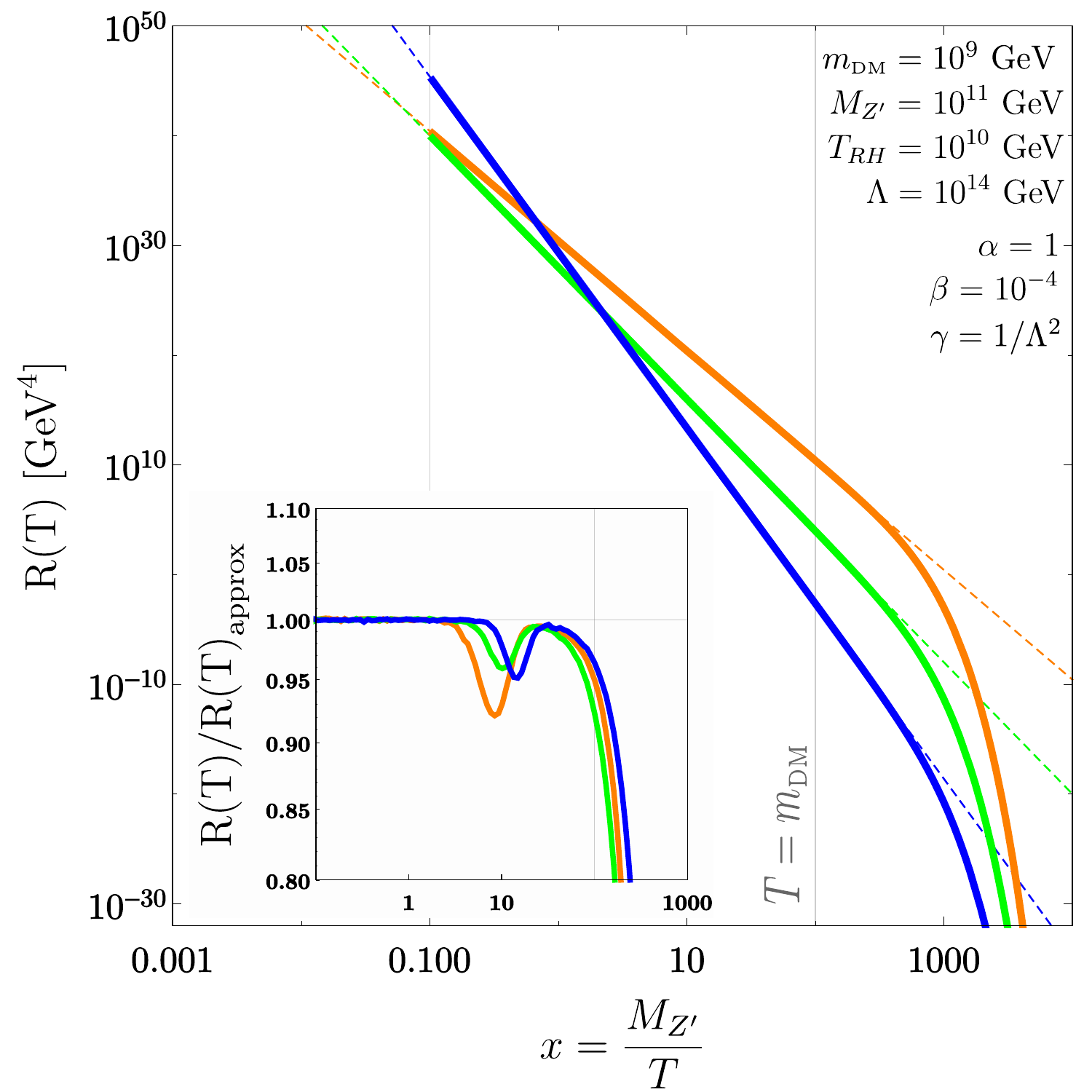}
\caption{\em \small Rates for the fermionic, Abelian and non-Abelian
  DM (orange, green and blue lines, respectively). Solid lines
  represent the exact numerical computation while dashed lines
  represent the approximated results based on Eqs.~(\ref{Eq:rate}). We
  fix $N_\psi = Q_\psi = 1$ and $m_\psi = 0.4 ~M_\zp$ and $T_\max=100
  ~T_\rh$ for illustrative purposes.  }
\label{Fig:rate}
\end{figure}

We have integrated numerically the production rate, Eq.~(\ref{Eq:rt}),
considering the Bose-Einstein distributions of the gluons and the
exact squared amplitudes of our three dark matter candidates. Our
result is depicted in Fig.\ref{Fig:rate}.

We can obtain analytical approximations for the rates by assuming
$\G_\zp \ll M_\zp$ and $m_\dm^2 \ll s$:
\begin{equation}\label{Eq:rate}
R(T) \approx \left\{
\begin{array}{rc} 
2 \times 10^2 ~ \dfrac{\alpha^2}{\Lambda^4} \dfrac{m_\c^2}{M_\zp^4}
T^{10} ~~~~ (\text{fermionic DM}) \\ \noalign{\medskip} 10^4 ~
\dfrac{\b^2} {\Lambda^4 M_\zp^4} T^{12} ~~~~~~ (\text{Abelian DM})
\\ \noalign{\medskip} 2 \times 10^9 \dfrac{\g^2}{\Lambda^4
  M_{Z^\prime}^4} T^{16} ~ (\text{non-Abelian DM})
\end{array} \right.
\end{equation}

We also show in Fig.~\ref{Fig:rate} our approximate solutions. In the
inset of the figure, we show when they depart from the exact
solutions. We can distinguish two regimes in which the approximations
fail. First, let us consider when the temperature of the thermal bath
is close to the mediator mass. In this case, the exact solutions are
smaller than the approximate results as an effect of the nonvanishing
mediator decay width. Even though the departure from approximations is
small in this case, it carries a special feature of our set-up,
emerging from the consequence of the Landau-Yang theorem in a thermal
bath of gluons.  The significant departure from approximations occurs
at large $x\equiv M_{Z'}/T$, due to a threshold effect, as for $T \ll
m_{\dm}$ the production rate is exponentially suppressed because only
the high-energy tail of the initial states distribution function have
sufficient energy to produce a DM pair, an effect that is not
encapsulated in the analytical approximations.

Another typical characteristic of a longitudinal (``would-be Goldstone") mediator appears in the generic expression for the
rate. Indeed, the ``light" mediator regime ($M_\zp \ll T_\rh$), and the
``heavy" mediator regime ($M_\zp \gg T_\rh$) give the same dependence
of the rate $R(T)$ on $M_\zp$, and thus on temperature for a given
nature of dark matter, as one can see from the Eq.~(\ref{Eq:rate}) and
Fig.~\ref{Fig:rate}. In fact, there exists only one main regime,
independent of the mass of the $Z'$ mediator\footnote{This is in
  contrast with what has been observed in \cite{mambrinilast} for
  spin-2 mediator.} for which the slope of the rate is constant until
$T\sim m_{\dm}$.

This can be understood by noting that only the longitudinal mode of
$Z'$ is exchanged, and hence it cannot feel any pole effect. The
longitudinal component has its origin in the Goldstone mode of a
nonlinear sigma model.  The behavior of the amplitude squared is
dominated by a term proportional to powers of $1/M_\zp$. This happens
because the Goldstone, which is the dominant mode exchanged in the DM
production process, carries the $1/V$ factor arising from $U(1)'$
breaking.  This is similar to the gravitino production in supergravity
in which the longitudinal mode, carrying a factor $1/m_{3/2}$ ($m_{3/2}$
being the gravitino mass), is generated in a high-scale supersymmetric
scenario as was shown in Ref.~\cite{gravitino}.

\vspace{-.5cm}
\subsection{Dark matter freeze-in}

For instantaneous reheating, the Universe is dominated by radiation
and entropy is conserved. In this case,
\begin{equation}
\frac{\text{d}}{\text{d}t} = - H(T) T
\frac{\text{d}}{\text{d}T}, \hspace{.5cm} \textrm{with}~~ H(T) =
\sqrt{\frac{g_e}{90}} \pi \frac{T^2}{M_P}
\end{equation}
and Eq.~(\ref{dndt}) can be put in the familiar form \beq
\frac{\text{d} Y}{\text{d}T} = - \frac{R(T)}{H~T~s} \,, \hspace{.5cm}
\textrm{for}~~ T < T_\rh
\label{Eq:y}
\eeq
where $Y=\frac{n}{s}$ is the dark matter yield, $s=\frac{2 \pi^2}{45}
g_s T^3$ is the entropy density of the thermal bath with $g_{e,s}$ the
energy and entropy density degrees of freedom, and $M_P \approx 2.4
\times 10^{18}$ GeV is the reduced Planck mass.

However, once we consider noninstantaneous reheating, dark matter can
be produced before the end of reheating. Indeed, as was shown in Refs.~\cite{Giudice:2000ex,Garcia:2017tuj}, a rate with a dependence $R(T)
\propto T^n$ for $n \geq 12$ enhances drastically the dark matter
production before $T_\rh$ if the reheating is considered as
noninstantaneous. For instance, for $n=12$, the ratio of the relic
abundance computed with the noninstantaneous reheating hypothesis
($\Omega h^2$) to the one with the instantaneous reheating hypothesis
($\Omega h^2_\rh$) is $\Omega h^2 /\Omega h^2_\rh \simeq 0.4\times
\dfrac{55}{6} \ln\left( \dfrac{T_\max}{T_\rh} \right)$ (where $T_\max$
is the maximum temperature produced in the reheating process), whereas
$\Omega h^2 /\Omega h^2_\rh \simeq 0.4 \times \dfrac{8}{5} \left(
\dfrac{n-5}{n-12} \right) \left( \dfrac{T_\max}{T_\rh} \right)^{n-12}$
for $n > 12$ and $\Omega h^2 /\Omega h^2_\rh \sim 2$ for $n=10$.
These ratios can be seen as ``boost factors" from the reheating
process and will be called $B^n_F \equiv \Omega h^2 /\Omega h^2_\rh$
from now on.  After integrating Eq.~(\ref{Eq:y}) from $T_\max$ until
the present day, we deduce the parameter space leading to the relic
abundance

\begin{widetext}
\begin{equation}\begin{split}
&\frac{\Omega h^2}{0.12} \approx \left\{
\begin{array}{lc} 
\left(\dfrac{B_F^{10}}{2}\right)\left(\dfrac{\a}{1}\right)^2 \left(
\dfrac{m_\chi}{6 \times 10^{10}~{\rm GeV}} \right)^3
\left(\dfrac{10^{14}~{\rm GeV}}{M_\zp} \right)^4
\left(\dfrac{10^{16}~{\rm GeV}}{\Lambda}\right)^4 \left(
\dfrac{T_\rh}{10^{12}~{\rm GeV}} \right)^5 \\ \noalign{\medskip}
\left(\dfrac{B_F^{12}}{21}\right)\left(\dfrac{\b}{10^{-5}}\right)^2
\left( \dfrac{m_\xa}{ 10^{9}~{\rm GeV}} \right)
\left(\dfrac{10^{14}~{\rm GeV}}{M_\zp} \right)^4
\left(\dfrac{10^{16}~{\rm GeV}}{\Lambda}\right)^4 \left(
\dfrac{T_\rh}{10^{13}~{\rm GeV}} \right)^7 \\ \noalign{\medskip}
\left(\dfrac{B_F^{16}}{1.76 \times 10^{8}}\right)\left(
\dfrac{m_\xna}{2 \times 10^{9}~{\rm GeV}} \right)
\left(\dfrac{10^{14}~{\rm GeV}}{M_\zp} \right)^4
\left(\dfrac{10^{16}~{\rm GeV}}{\Lambda}\right)^8 \left(
\dfrac{T_\rh}{10^{12}~{\rm GeV}} \right)^{11}
\end{array} \right.
\label{Eq:omega}
\end{split}\end{equation}
\end{widetext}
where we set $\gamma= 1/\Lambda^2$, and we consider $SU(2)$ as gauge
group for the non-Abelian dark matter case.

\begin{figure}[ht!]
\centering
\includegraphics[scale=0.6]{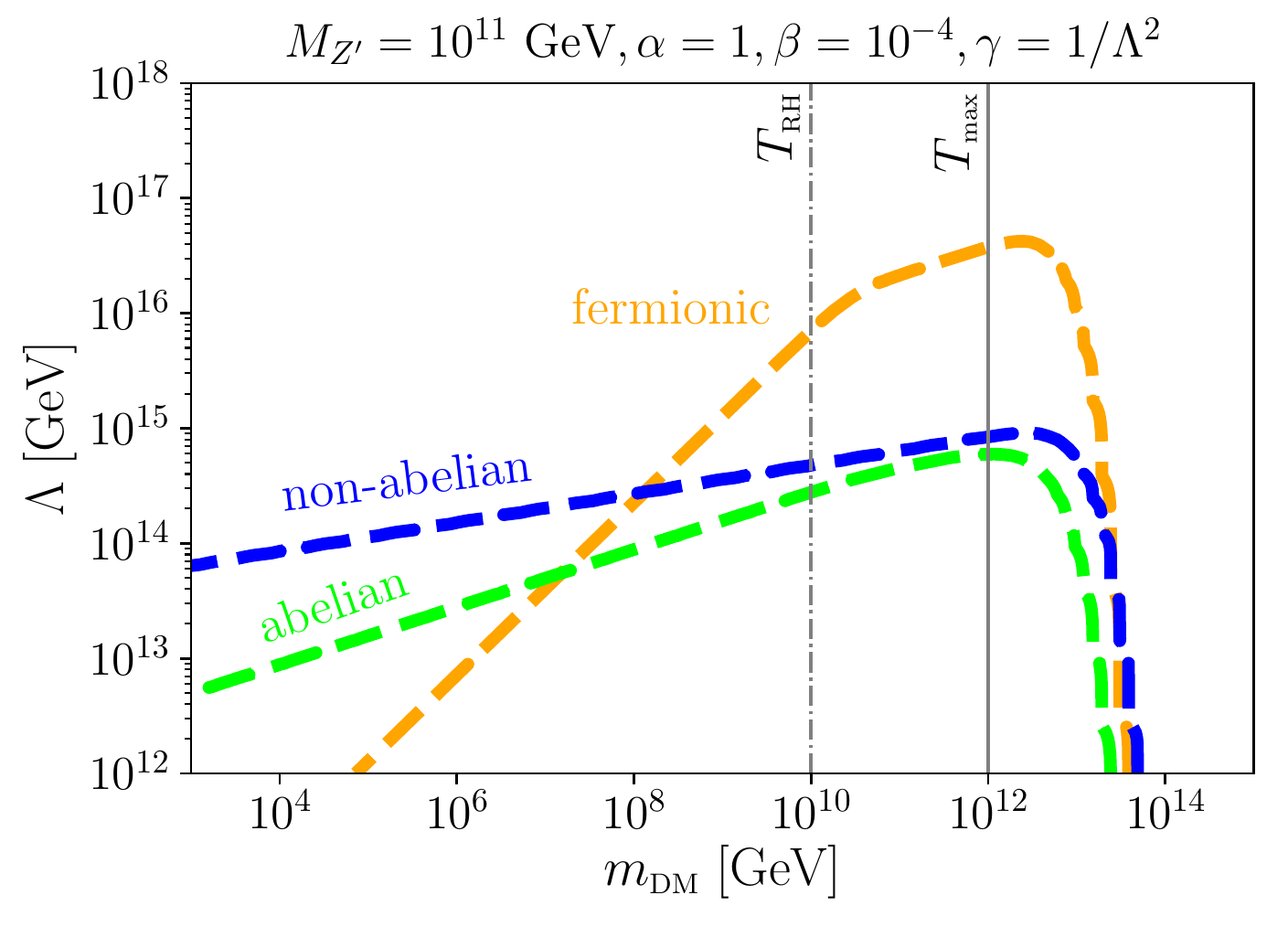}
\caption{\em \small Relic density curves for the fermionic (orange
  dashed line), Abelian (green dashed line) and non-Abelian (blue
  dashed line) dark matter.}
\label{effectiveapproach}
\end{figure}

Our results are summarized in Fig.~\ref{effectiveapproach}, in which we
plotted the parameter space allowed by the cosmological constraints in
the plane $(m_\dm,\L)$ for the three natures of dark matter considered
in this work and for $T_\rh = 10^{10}~\text{GeV}$ and $T_\max = 100~
T_\rh$.  To compute the relic abundance numerically, we did not assume
instantaneous reheating. Instead, using the method developed in Ref.~\cite{Garcia:2017tuj}, we assumed that an inflaton $\phi$ decays into
radiation with a rate $\Gamma_\phi$ and that the DM particles are
created and annihilated into radiation of density $\rho_{\text R}$
with a thermal-averaged cross section times velocity $\langle\sigma
v\rangle$.  The corresponding energy and number densities satisfy the
differential equations~\cite{Chung:1998rq, Giudice:2000ex} 
\bea
\frac{\text{d}n_\dm}{\text{d}t}&=&-3H\,n_\dm-\langle\sigma
v\rangle\left[n_\dm^2-(n_\dm^\text{eq})^2\right]\,, \nonumber
\\ \frac{\text{d}\rho_{\text R}}{\text{d}t}&=&-4H\,\rho_{\text
  R}+\Gamma_\phi\,\rho_\phi+2\langle\sigma v\rangle\langle
E_\dm\rangle\left[n_\dm^2-(n_\dm^\text{eq})^2\right]\,, \nonumber
\\ \frac{\text{d}\rho_\phi}{\text{d}t}&=&-3H\,\rho_\phi-\Gamma_\phi\,\rho_\phi\,
,
\label{Eq:setboltzmann}
\eea with $\langle E_\dm \rangle$ the mean energy of the dark matter.
The Hubble expansion parameter $H$ is given by
$H^2=\frac{1}{3M_P^2}(\rho_\phi+\rho_{\text R}+\rho_\dm)$, and its
dependence in temperature is quite complex between $T_\max$ and
$T_\rh$ because of the composition of the Universe (mixed between a
decaying inflaton compensated by an increasing radiation\footnote{We
  have defined $T_\rh$ as the temperature in a radiation-dominated
  Universe after the inflaton decay, $\Gamma_\phi = H(T_\rh)$}).

We can see that our analytical expressions in Eq.~(\ref{Eq:omega})
give an impressively good approximation. Indeed, the right amount of
relic abundance \cite{planck} is obtained (in the fermionic case) for
$m_\chi=10^6$ GeV, $T_\rh=10^{10}$ GeV, $M_\zp=10^{11}$ GeV, and
$\Lambda\simeq 8\times 10^{12}$ GeV, in perfect agreement with
Fig.~\ref{effectiveapproach}.  The slopes of the curves depicted in
Fig.~\ref{effectiveapproach} correspond also perfectly with the ones
predicted by our analytical solution in Eq.~(\ref{Eq:omega}): it
follows a line $\Lambda \propto m_\dm^{3/4}$ ($m_\dm^{1/4}$,
$m_\dm^{1/8}$) for fermionic (Abelian, non-Abelian) for $m_\dm <
T_\max$.

Without entering too much into detail, there is an interesting
feature in the change of slope between $T_\rh$ and $T_\max$ in the
fermionic dark matter case.  This is a novel feature that was not
treated in Ref.~\cite{Garcia:2017tuj} nor Ref.~\cite{mambrinilast}. Indeed, in
the case in which dark matter is heavier than $T_\rh$, there is still a
possibility to produce it as long as $m_\dm \lesssim T_\max$. If the
temperature dependence of the rate is small enough (fermionic case),
most of the DM density is produced at the lowest scale available, and
we notice a change of slope in the curve giving the correct relic
density.  It is worth commenting that, due to statistical
distribution, the production rate does not vanish completely when $T
\lesssim m_\dm$, which explains why the DM production window is still
open when $m_\dm> T_\max$\footnote{The corresponding region of
  parameter space as shown in Fig.~\ref{effectiveapproach} is
  quantitatively less precise as the EFT approach becomes less
  reliable.}. Therefore, in this regime, a small effective scale
$\Lambda$ is required to compensate the thermal suppression of the rate, as one can see in Fig.\ref{effectiveapproach}.

Moreover, a quick look at Fig.\ref{effectiveapproach} shows to what
extent the allowed parameter space is technically natural.  Indeed,
for a very large range of the DM mass, from $\mathcal{O}$(TeV) to
$T_\rh$, values of the beyond SM scale $\L$ range from $T_\rh$ to
GUT/string scale and can still populate the Universe with the correct
relic abundance.  This means that the heavy spectrum of masses above the
reheating temperature $T_\rh$ generates naturally small couplings of
an invisible $Z'$ to the SM bath to satisfy the cosmological
constraints through the freeze-in process. This constitutes one of the
most important observations of our work.

\section{IV. toward a microscopic approach}

\begin{figure}[t]
\centering
\includegraphics[scale=0.29]{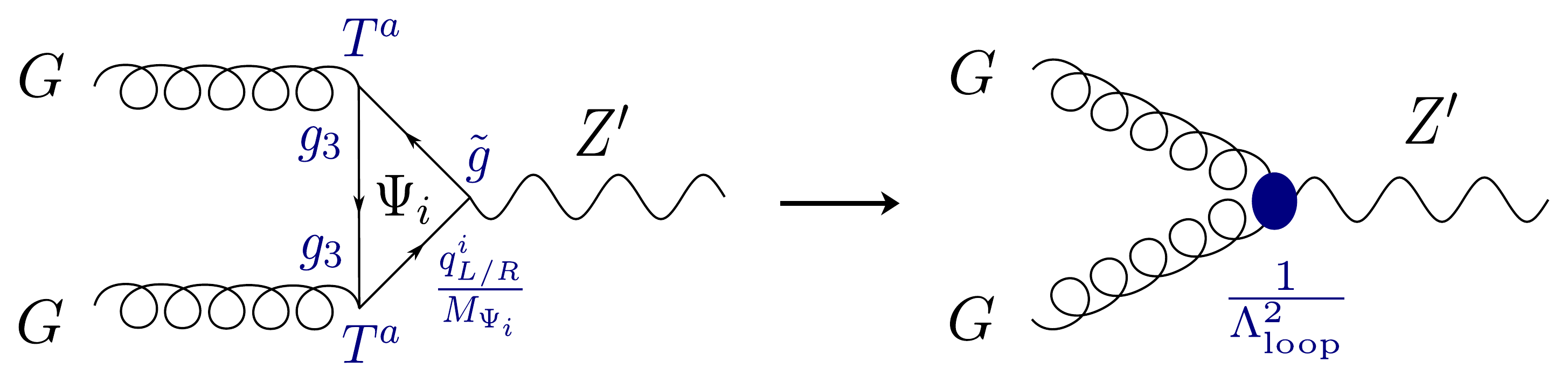}
\caption{\em \small Triangle diagram containing heavy chiral fermions $\Psi_i$ (left panel) and the resulting effective vertex at
  low energy (right panel).}
\label{Fig:loops}
\end{figure}

As mentioned earlier, we consider processes happening at a temperature
below the $U(1)'$ phase transition scale.  We have also assumed that
the radial component of the complex scalar that breaks $U(1)'$ is way
too heavy compared to the corresponding VEV ($V$). Then, $Z'$ is
primarily longitudinal absorbing the axion field ($a$), and the
effective Lagrangian containing $Z'$ realizes the gauge symmetry
nonlinearly \`a la Stueckelberg. Now, we attempt to look deep inside
the effective GCS vertices searching for microscopic
details. Importantly, the masses of the loop fermions ($\Psi$)
generating the GCS couplings, as shown in Fig.~\ref{Fig:loops}, must
be invariant both under the SM and the $U(1)'$ gauge symmetries to
ensure that the induced low-energy GCS operators are gauge
invariant. One can, in fact, write the microscopic (gauge-invariant)
Lagrangian introducing pairs of heavy fermions ($\Psi$) that are
vectorlike with respect to the SM group, but necessarily chiral under
$U(1)'$. This generates the effective Lagrangian
(\ref{Eq:leff}) at energies below the $U(1)'$ breaking scale,
\begin{equation}\label{Eq:microscopic}
\begin{split}
{\cal L} =& {\cal L}_\text{SM} + \frac{1}{2}(\partial_\mu a - M_{Z'}
Z'_\mu)^2 - M_i~\overline{\Psi}^i_L e^{i(q_L-q_R)\frac{a}{V}} \Psi^i_R
\\ & \hspace{-.2cm} + i \overline{\Psi}^i_L \gamma^\mu (\partial_\mu
-i \frac{\tilde g}{2} q_L^i Z'_\mu) \Psi^i_L + i \overline{\Psi}^i_R
\gamma^\mu (\partial_\mu -i \frac{\tilde g}{2} q_R^i Z'_\mu) \Psi^i_R
\end{split}
\end{equation}
which is manifestly invariant under the (nonlinear) $U(1)'$
transformation of parameter $\alpha$, \bea && \Psi^i_{R} \rightarrow
\Psi^i_{R} e^{i \frac{\tilde g}{2}q_{R} \alpha }~; ~~~~ \Psi^i_{L}
\rightarrow \Psi^i_{L} e^{i \frac{\tilde g}{2}q_{L} \alpha } \nonumber
\\ && Z'_\mu \rightarrow Z'_\mu + \partial_\mu \alpha ~; ~~~~ a
\rightarrow a+ \frac{\tilde g}{2} V ~ \alpha \equiv a + M_{Z'} ~
\alpha \nonumber \eea

From the Lagrangian in Eq.~(\ref{Eq:microscopic}), we compute the
triangle loops shown in Fig.~\ref{Fig:loops} and integrate out the
heavy fermions. We then obtain the same effective Lagrangian as in 
Eq.~(\ref{Eq:leff}), but now we can express the effective coupling of
the dimension-6 Lagrangian in terms of the parameters of the
microscopical theory. In agreement with Ref.~\cite{Dudas:2013sia}, we obtain

\beq
{\cal L}_\text{loop}= \frac{1}{\Lambda_\text{loop}^2}
\partial^\alpha Z'_\alpha \epsilon^{\mu \nu \rho \sigma}
\mathrm{Tr} [G^a_{\mu \nu} G^a_{\rho \sigma}],
\eeq
with 
\beq
\frac{1}{\Lambda_\text{loop}^2} = \frac{g_3^2~ \tilde g}{96 \pi^2}
\sum_i \frac{q_L^i - q_R^i}{M_{\Psi_i}^2} \mathrm{Tr}[T^a T^a].
\label{eq:lambdaloop}
\eeq

Defining for simplicity $\sum_i \frac{q_L^i - q_R^i}{M_{\Psi_i}^2}
\mathrm{Tr}[T^a T^a] = \frac{N_\Psi Q_\Psi}{M_\Psi^2}$ (which
corresponds to a set of $N_\Psi$ fermions of effective charges
$Q_\Psi$ and masses $M_\Psi$) we obtain $\Lambda_\text{loop} \simeq
\frac{50}{\sqrt{N_\Psi Q_\Psi}} \frac{M_\Psi}{\sqrt{\tilde
    g}}$\footnote{We take the SM expected value of $g_3$ at
  $10^{12}~\text{GeV}$.}. We can now reexpress the production rates in
Eqs.~(\ref{Eq:rate}) in terms of the fundamental parameters of the
microscopic theory. For the fermionic dark matter case, we then have
\beq
R(T) \simeq 5 \times 10^{-4} 
\left(\frac{\alpha ~N_\Psi Q_\Psi}{y_\Psi^2 \tilde g}\right)^2
\frac{m_\chi^2}{V^8} T^{10}
\eeq
where we defined $M_\Psi = y_\Psi V$ and $M_{Z'}= \frac{\tilde g}{2}
V$. Solving Eq.~(\ref{Eq:y}) gives (with all mass dimensional
parameters in GeV units) 
\beq \dfrac{\Omega h^2}{0.12} \simeq \left(
\dfrac{B_F^{10}}{2} \right) \left(\frac{\alpha N_\Psi Q_\Psi}{\tilde g
  y_\Psi^2}\right)^2 \left(\frac{m_\chi}{10^{10}}\right)^3 \left(
\frac{T_\text{RH}}{10^{12}} \right)^5 \left( \frac{10^{14}}{V}
\right)^8 .
\label{Eq:omegauv}
\eeq

We could keep $V$ as a free fundamental parameter of the model, which
is determined by the potential of the Higgs responsible for the extra
$U(1)'$ breaking. However, to be more complete, we investigated UV
scenarios in which $V$ is determined as an intermediate scale by the
unification condition of the gauge coupling constants, in $SO(10)$ GUT
constructions (as an example).

\begin{figure}[h!]
\includegraphics[scale=0.6]{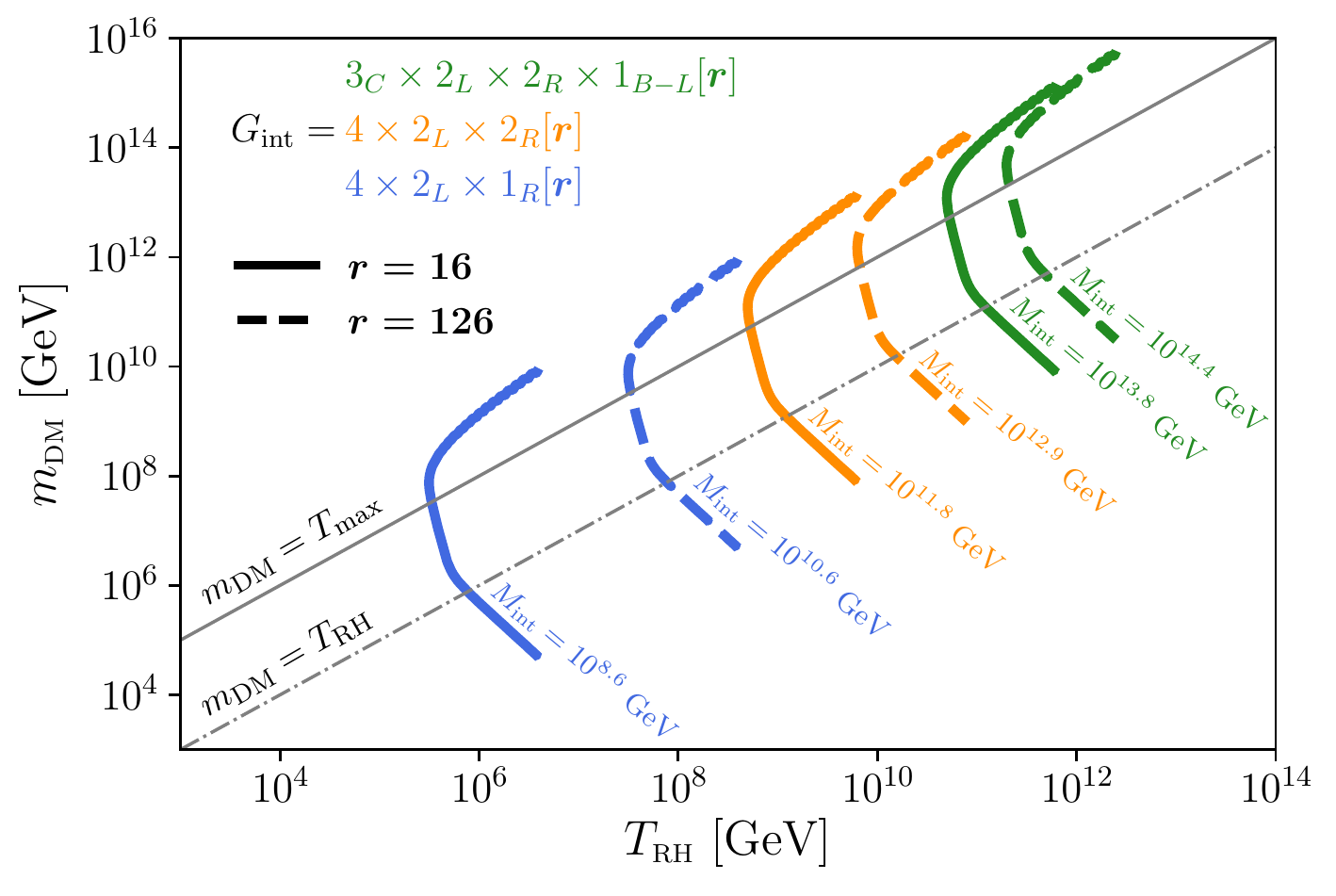}
\caption{\em \small Relic density curves for the fermionic DM case in
  several $SO(10)$ breaking schemes and for different representations
  \textnormal{[$\boldmath{r}$]} for the Higgs field responsible for
  intermediate scale breaking. We take $N_\Psi = Q_\Psi = y_\Psi = 1$
  for illustrative purposes.}
\label{Fig:uv}
\end{figure}

Indeed, in such set-ups, the $SO(10)$ group is not directly broken
into the SM in one step but goes through an intermediate gauge group
$G_\text{int}$ like $SO(10) \rightarrow G_\text{int} \rightarrow
SU(3)_c \times SU(2)_L \times U(1)_Y$. The scale $M_\text{int}$ at
which the intermediate gauge group is broken is fixed by the
unification condition $g_1=g_2=g_3$ at a higher unified scale. It was
shown in Refs.~\cite{Mambrini:2013iaa} (at one loop) and~\cite{2loops} (at
two loops) that $V = M_{\rm int}$ can range from $10^9$ to
$10^{15}$ GeV depending on $G_\text{int}$ and the representation in
which the Higgs field responsible for $G_\text{int}$ breaking lies.
We show in Fig.~\ref{Fig:uv} the parameter space providing the correct
relic density for our fermionic dark matter candidate in several
intermediate scenarios. Here, we take $V=M_\Psi=M_\text{int}$, which is
a reasonable approximation. The numerical results, obtained by solving
the complete set of Boltzmann equations, Eq.~(\ref{Eq:setboltzmann}),
and numerical integration of the rate, Eq.~(\ref{Eq:rt}), are in
perfect agreement with our analytical solution Eq.~(\ref{Eq:omegauv}).

We observe that in these unified scenarios DM density corresponding
to the Planck measurements~\cite{planck} can be directly produced from
annihilation of SM particles even if the mediator $Z'$ is extremely
heavy with no SM particles charged under the extra $U(1)'$. The
effective couplings of the thermal bath to $Z'$ are being generated
through the GCS interactions.

Figure~\ref{Fig:uv} highlights the natural relation between the
parameters of the theory $M_{\text{int}}\sim T_\max \sim 10^2 ~T_\rh
\sim 10^2 ~m_\dm$ corresponding to the correct DM relic
abundance. Therefore, the intermediate scale in such unified
constructions could be closely related to the inflaton mass as one
expects it to be of the order of the maximum temperature reached by the
SM thermal bath. The large hierarchy between these scales and the SM
electroweak VEV naturally provides the suppressed DM-SM effective
coupling required to produce the correct DM density nonthermally via
the freeze-in mechanism.

Though the heavy colored fermions $\Psi$ are cosmologically stable,
their very small number density owing to their heaviness ($\sim
10^{14}$ GeV) keeps them hidden as benign\footnote{For a discussion of
  the possibility of the DM as a colored composite object of mass
  $\sim 10$ TeV reproducing the relic abundance, see a recent
  analysis \cite{DeLuca:2018mzn}, and also an older one
  \cite{Smith:1982qu}.}. In principle, these states contribute to the
running of $g_3$, but they are kicked into life so late that the
nonstandard modification of running during the remaining phase up to
the GUT scale is inconsequential.

If, however, the $\Psi$ states additionally have SM hypercharge, there
is a nontrivial twist. Through loop contributions, they would induce a
kinetic mixing term like $\delta B^{\mu \nu} Z^\prime_{\mu \nu}$,
which would ascribe the kinetically diagonal $Z'$ with a small,
proportional to $\delta$, coupling with the SM fermion $f$. This would
allow the direct s-channel production of DM from $f$ annihilation as
follows: $\bar{f}+f\rightarrow Z^\prime \rightarrow
\text{DM}+\text{DM}$. Note that the natural size of $\delta$ is a loop
factor times a logarithm of the ratio of two scales, so $\delta \sim
10^{-4}$ is a representative number. If $M_{Z'} < T_{\rm MAX}$, then
through $Z'$, the DM would obviously be in thermal equilibrium with the
SM. But this does not constitute the FIMP or freeze-in scenario we are
pursuing here. On the other hand, if $M_{Z'} > T_{\rm MAX}$, the above-mentioned tree-level $s$-channel annihilation, as shown in
\cite{Mambrini:2013iaa}, would be the dominating process.  The
GCS coupling-induced process we advocated in this paper would then be
a subleading one. The importance of our analysis lies in the fact
that even if the heavy $\Psi$ states do not carry the SM hypercharge
and the kinetic mixing is absent, the novel mechanism triggered by the
GCS interaction can still explain the relic DM abundance  {\em albeit}
for an inaccessibly high-mass range of both the DM and its portals.

\section{V. Conclusion}
We have shown that a dark (very) massive $Z'$, not charged under the
SM gauge group, can successfully play the role of a mediator between
the visible and the dark sectors even if the corresponding $U(1)'$
breaking scale lies far above the maximum temperature of the
Universe.  Pair annihilation of SM gauge bosons, proceeding through
triangle loops containing heavy fermions through this $Z'$ portal, can
produce a cosmologically agreeable amount of DM.  These types of
effective couplings between the SM gauge bosons and $Z'$ find inherent
justification in an anomaly-free set-up in which Chern-Simons (more
precisely, the GCS type discussed in the beginning) terms are
generated through the anomaly cancellation mechanism.

The large effective scale of the GCS operators is responsible for the
weakness of the DM-SM interaction strength without invoking
unnaturally small couplings as is often required in the context of
freeze-in. The large dependence of the production rate on temperature
indicates that the majority of DM is produced during the initial
moments of reheating. Subsequently, the reheating process itself lends
important consequences to the computation of DM production. Moreover,
the assertion that the $U(1)'$ breaking scale lies far above the
reheating temperature implies that only the longitudinal mode (axion
{\em \`a la} the Stueckelberg formalism) of the $Z'$ contributes to the
production process, rendering its phenomenology very particular in
comparison with other type of mediators. The prominence of the
longitudinal mode is also consistent with how $Z'$ is coupled via GCS
interaction. Such a scenario can be embedded in a unified $SO(10)$
framework in which the $Z'$ mass scale represents an intermediate
breaking stage of $SO(10)$.

\vskip.1in
{\bf Acknowledgments:}
\noindent 
The authors want to thank especially E. Dudas for very insightful
discussions. This research has been supported by the (Indo-French)
CEFIPRA/IFCPAR Project No.~5404-2. Support from CNRS LIA-THEP and the
INFRE-HEPNET of CEFIPRA/IFCPAR is also acknowledged. G.B. acknowledges
support of the J.C.~Bose National Fellowship from the Department of
Science and Technology, Government of India (SERB Grant
No.~SB/S2/JCB-062/2016). M.D. acknowledges support from the Brazilian
Ph.D. program ``Ci\^encias sem Fronteiras'' -- CNPQ Process
No. 202055/2015-9. This work was also supported by the France-US PICS
no. 06482, PICS MicroDark. M. D. and Y. M. acknowledge
partial support from the European Union Horizon 2020 research and
innovation programme under the Marie Sklodowska-Curie: RISE
InvisiblesPlus (Grant No 690575) and the ITN Elusives (Grant No 674896).

\section*{Appendix}
\section{noninstantaneous reheating: analytical estimations}

As the reheating is not an instantaneous process, the inflaton
dominates the energy density of the Universe at some stage and we have
a different relation between time and temperature
\cite{Giudice:2000ex} \footnote{The numerical factors $c$ and $d$ are
  related to the convention chosen in the definition of reheating
  temperature: $t_\phi = \frac{c}{\Gamma_\phi}$ is the time of
  inflaton decay completion and $t_H = \frac{d}{H}$ is the Hubble
  time. The reheating temperature is defined such that $\Gamma_\phi =
  \frac{c}{d}H(T_\rh)$.}
\begin{equation}\label{inflatonera}
\frac{\text{d}}{\text{d}t} = - \frac{3}{8} H(T) T
\frac{\text{d}}{\text{d}T}, ~~ \textrm{with} ~~ H(T) =
\frac{d}{c}\sqrt{\frac{5 g_e^2}{72 g_\rh}} \pi \frac{T^4}{T_\rh^2
  M_P},
\end{equation}
with $g_\rh = g_e(T_\rh)$.

We emphasize that while the inflaton dominates, we have $T \propto
a^{-3/8}$ instead of $a^{-1}$ as in the radiation-dominated era, with
$a$ the expansion scale factor.

It is convenient to define the dimensionless quantity
$\mathcal{N}_\phi \equiv \rho_\phi/T_\rh a^3$ \cite{Giudice:2000ex},
with $\rho_\phi$ the energy density of the inflaton field. It remains
constant before the end of reheating, with a value given
by \footnote{We have $\mathcal{N}_\phi^I \equiv b_2^{-8}
  \frac{g_\max^2}{g_\rh}(\frac{T_\max}{T_\rh})^8$, with $b_2 =
  \Big(\frac{3^{11/10}5^{1/2}}{2^{23/10}\pi}\frac{c}{d} \Big)^{1/4}$
  and $g_\max = g_e(T_\max)$.} $\mathcal{N}_\phi^I \propto
(\frac{T_\max}{T_\rh})^8$.

We notice that, by using Eq.~(\ref{inflatonera}), Eq.~(\ref{dndt}) can
be put in the form
\begin{equation}\label{yid} 
\frac{\text{d}Y_\id}{\text{d}T} = - \frac{8}{3}
\frac{R(T)}{H~T~\epsilon}, \hspace{.5cm} \textrm{for}~~ T_\rh < T <
T_\max
\end{equation}
where the yield in the inflaton-dominated era is defined as the
dimensionless and comoving parameter $Y_\id \equiv n/\epsilon$ and
$\epsilon$ is defined as
\begin{equation}
\epsilon \equiv \mathcal{N}_\phi^I a^{-3} = \frac{5 d^2 g_e^2}{96 c^2 g_\rh T_\rh^5} T^8.
\end{equation}

By solving Eq.~(\ref{yid}), we can find the contribution of the
noninstantaneous heating process to the relic density and define the
boost factor discussed in Sec. III.

\section{Decay width}

We present the decay width used in the computation of the production
rate. Allowing the $Z'$ to decay into $N_\psi$ dark heavy fermions
$\psi$ (with only vectorial coupling) of charge $Q_\psi$ and mass
$m_\psi$, the total decay width is given by

\begin{equation}\label{width}
\Gamma_{Z'} = \frac{M_\zp}{12\pi}N_\psi Q_\psi^2 \sqrt{1-\frac{4
    m_\psi^2}{M_\zp^2}}\Big(1+\frac{2m_\psi^2}{M_\zp^2}\Big) +
\Gamma_{\zp \to \text{DM}},
\end{equation}
where
\begin{equation}
\Gamma_{\zp \to \text{DM}} = \frac{M_\zp}{12\pi}\times\left\{
\begin{array}{rc} 
\a^2 \Big(1-\frac{4m_\chi^2}{M_\zp^2}\Big)^{3/2} ~ (\text{fermionic
  DM}) \\ \noalign{\medskip} \frac{\b^2}{2} \frac{M_\zp^2}{m_\xa^2}
\Big(1-\frac{4m_\xa^2}{M_\zp^2}\Big)^{5/2} ~~~ (\text{Abelian DM})
\\ \noalign{\medskip} 0 ~~~~~~~~~ (\text{non-Abelian DM})
\end{array} \right.
\end{equation}

The decay width of $Z'$ into non-Abelian DM vanishes identically
because of the nature of the coupling.

\vspace{-.5cm}
\bibliographystyle{apsrev4-1}

\end{document}